\documentclass[universe,article,accept,pdftex,moreauthors]{Definitions/mdpi} 

\firstpage{1} 
\pubvolume{1}
\issuenum{1}
\articlenumber{0}
\pubyear{2025}
\copyrightyear{2025}
\externaleditor{Firstname Lastname} 
\datereceived{4 February 2025 } 
\daterevised{3 March 2025 } 
\dateaccepted{5 March 2025 } 
\datepublished{ } 
\hreflink{https://doi.org/}

\usepackage{bm}

\Title{A Hubble Constant Determination Through Quasar Time Delays and Type Ia Supernovae}

\TitleCitation{A Hubble Constant Determination Through Quasar Time Delays and Type Ia Supernovae}

\Author{Leonardo R. Colaço}

\AuthorNames{Leonardo R. Colaço}

       \AuthorCitation{Colaço, L.R.}

\address[1]{Departamento de F\'{i}sica Te\'{o}rica e Experimental, Universidade Federal do Rio Grande do Norte, 59300-000 Natal,  RN, Brazil; colacolrc@gmail.com\\
}

\abstract{This paper presents a new model-independent constraint on the Hubble constant ($H_0$) by anchoring relative distances from Type Ia supernovae (SNe Ia) observations to absolute distance measurements from time-delay strong Gravitational Lensing (SGL) systems.	 The approach only uses the validity of the cosmic distance duality relation (CDDR) to derive constraints on $H_0$. By using Gaussian Process (GP) regression to reconstruct the unanchored luminosity distance from the Pantheon$+$ compilation to match the time-delay angular diameter distance at the redshift of the lenses, one yields a value of \mbox{$H_0 = 75.57 \pm 4.415$ km/s/Mpc} at a 68\% confidence level. The result aligns well with the local estimate from Cepheid variables within the $1\sigma$ confidence region, indicating consistency with late-universe probes.}


\begin{document}

\section{Introduction}

The best way to interpret the nature of cosmology is through a theoretical framework that incorporates the coexistence of Cold Dark Matter (CDM) and Dark Energy (DE) in the absence of geometrical curvature.~CDM is non-relativistic, while DE, characterized by the constant $\Lambda$, is responsible for the current expansion phase. This framework is known as the flat $\Lambda$CDM model. It effectively describes the structure and evolution of the universe, owing to its relative simplicity and capability to predict the current expansion phase~\cite{Perlmutter1999,Riess1998}. However, this model faces several issues widely discussed within the scientific community, raising concerns about its reliability. The~most significant issue is known as the Hubble constant tension (or simply $H_0$-tension), which refers to a notable disagreement between independent early (global) and late (local) time measurements of $H_0$. Local measurements use methods like the distance ladder, which extends to Type Ia Supernovae (SNe Ia) \cite{Riess2016jrr,Riess2021jrx,Reid2019}, calibrated through geometric distances, Cepheid variables, the~tip of the red giant branch (TRGB) \cite{Freedman2021,ScolnicL31}, and~using gravitational wave (GW) measurements~\cite{Abbott2018}. In contrast, global measurements rely on the assumption of a standard cosmological model and encompass observations of large-scale structures, such as Baryon Acoustic Oscillations~\cite{Pogosian2020, Alam2021} and~the so-called Cosmic Microwave Background (CMB) \cite{Planck2020}. Currently, the~$H_0$ tension between these measurements ranges from $4$ to $6$ $\sigma$. Several researchers have proposed alternative theories and methods to address this tension, but~a solution has yet to be found (see~\cite{Perivolaropoulos2024yxv,Abdalla2022yfr,DiValentino2021izs,Hu2023jqc,McGaugh} for reviews).

$H_0$-tension has also prompted new questions regarding the foundations of basic cosmological principles~\cite{Krishnan2021dyb}. Numerous proposals have been put forth to address this tension, but~their impact seems to diminish with every new set of data released~\cite{DiValentino2021izs,Beenakker2021vff}. One approach to resolving $H_0$-tension involves exploring various parametric models of the supernova absolute magnitude, expressed as $M_B = M_B(z)$. For~example, a~recent study from~\cite{Benisty2022psx} investigates the reliability of $M_B$ using non-parametric reconstruction techniques. In~this context, the~degeneracy between $H_0$ and $M_B$ is transformed into a degeneracy between $M_B$ and the sound horizon at the drag epoch $r_d$. Through Bayesian analysis and considering BAO measurements, the~authors evaluated different phenomenological models and found no strong preference for any specific model. In~another study from~\cite{Alestas2021luu}, the~authors also examined $M_B$ through a class of models that propose gravitational transitions at low redshifts ($z_t<0.1$) as a solution to the Hubble tension. By~using the complete CMB temperature anisotropy spectrum data, along with BAO and Pantheon SNe Ia datasets, the~authors found significant statistical advantages in addressing the Hubble tension with this class of models compared with smooth $H(z)$ deformation models ($\omega$CDM, CPL, and PEDE) (see~\cite{Perivolaropoulos2021bds,Mortsell2021nzg} for more information and the references therein).

A powerful technique known as strong lensing time-delay cosmography (TDC) provides a direct measurement of $H_0$ that does not rely on the local distance ladder or methods anchored to sound horizon physics~\cite{1964MNRAS128307R,Birrer2022chj,Saha2024axf}. Recent advancements in photometric precision have led to the discovery of many multiply imaged quasars and supernovae~\cite{2018MNRAS4811041T,2023MNRAS5203305L,Rodney2021keu}, along with improved time-delay measurements~\cite{Millon2020ugy,2017arXiv170609424C}. The~$H_0$ lenses in the COSMOGRAIL Wellspring (H0LiCOW) collaboration recently reported a value of \mbox{$H_0 = 73.2 \pm 1.75$ km/s/Mpc} achieving a precision of $2.4\%$ \cite{Wong2019}. This precision was refined to $2\%$ by incorporating blind measurements~\cite{Millon2019slk}. The~collaboration achieved a high accuracy by modeling lens galaxies with either a power law or a combined stars and dark matter mass profile. When these assumptions were relaxed, the~precision decreased to $8\%$, consistent with the results from both {\bf \textit{Planck}} 
 and SH0ES~\cite{Birrer2020tax}.

The model-independent determination of $H_0$ is particularly significant. Recently, a~method called the inverse distance ladder has been proposed~\cite{Aubourg15,Cuesta3463}, where the basic idea behind this method is to anchor the relative distances from Type Ia supernovae (SNe Ia) with the absolute distance measurements from other cosmological methods. Following this proposal, several works have been published (see~\cite{Liao2020zko,Liao2019qoc,Li2024elb,Li2024103,Gong2024yne} and the references therein). For~instance, by~combining the Pantheon sample with measurements from galaxy clusters (GCs) under minimal cosmological assumptions, the~authors in~\cite{Colaco2023gzy} reported a value of $H_0 =  67.22 \pm 6.07$ km/s/Mpc at $1\sigma$ confidence level, with~an error of approximately $9\%$. Using two galaxy cluster gas mass fraction measurement samples, Type Ia supernovae luminosity distances, and~the validity of the cosmic distance duality relation, the~authors from~\cite{Gonzalez2024qjs} provided an estimate of $H_0$, which is also independent of any specific cosmological framework. Their joint analysis yielded $H_0 = 73.4 \pm 5.95$ km/s/Mpc at a $68\%$ confidence level. Moreover, in~Ref.~\cite{Camarena2019rmj}, the~authors present a novel method for constructing the inverse distance ladder by linking local astrophysical measurements to the CMB. They employed CDDR, ensuring their results remain independent of any specific parameterization of the luminosity-distance relation and cosmological model. By~incorporating the latest data of SNe Ia, baryon acoustic oscillations (BAOs), and~CMB, the~authors achieved a consistently low value for the parameter $q_0$. They identified a notable inconsistency between constraints obtained from angular-only BAO measurements and those derived from anisotropic BAO~measurements. 

This paper aims to constrain the Hubble constant in a cosmological model-independent way by anchoring relative luminosity distances from SNe Ia with time-delay angular diameter distances of strong lensing systems. To~accomplish this, one uses the most extensive compiled dataset of SNe Ia to date, known as Pantheon$+$ \cite{Scolnic2021amr}, and~two datasets of two-image time-delay lensing systems~\cite{2013MNRAS4311528B,Wong2019}. This paper is organized as follows: Section~\ref{Methodology} introduces the methodology employed and the datasets used to perform the statistical analyses. The~findings regarding the constraints on $H_0$ are addressed in Section~\ref{rasults}, followed by the final remarks in Section~\ref{conclusions}.

\section{Materials and~Methods}\label{Methodology}

The primary method for constraining the Hubble constant is based on the effectiveness of the cosmic distance duality relation (CDDR), which is expressed as  $D_L = (1+z)^2 D_A(z)$. In~this equation, $D_L$ represents the luminosity distance, $D_A$ denotes the angular diameter distance, and~$z$ is the redshift. This relationship holds when the number of photons is conserved along null geodesics in a Riemannian spacetime between the observer and the source. Recent technological advancements have provided numerous observational data that enable a detailed examination of any potential deviations from the CDDR. However, no strong statistical evidence for such deviations has been identified up to this moment (see~\cite{Wang2024rxm,Favale2024sdq,Kumar2021djt,Lima2021slf,Li2023frp,Jesus2024nrl} and the references therein). The~broad applicability of this relation emphasizes its fundamental significance in observational cosmology, and~any violation of it could indicate new physics or systematic errors in the observational data~\cite{Ellis2007,CDDR}. Following the philosophical approach outlined in~\cite{Renzi2020fnx}, one can express CDDR further as
\begin{equation} \label{DLNA}
    H_0 = \frac{1}{(1+z)^2} \frac{\Theta^{\textrm{SNe}}(z)}{D_A(z)},
\end{equation}
where $ \Theta^{\textrm{SNe}}(z) \equiv [H_0 D_L(z)]^{\textrm{SNe}}$ represents the unanchored luminosity distance. This relationship highlights the potential for deriving estimates of $H_0$ if one has measurements of both the unanchored luminosity distance and the angular diameter distance at the same redshift $z$. Furthermore, this relationship is based on the assumption that the validity of CDDR is equivalent to the idea that all distance probes consistently trace cosmic expansion. If~this assumption were invalid, the~value of $H_0$ from Equation~(\ref{DLNA}) would not remain constant; instead, it would exhibit an unphysical trend with redshift. It highlights the role of $ H_0 $ not only as an absolute distance scale but also as a factor that exposes inconsistencies among distance~probes.

This paper takes the $\Theta^{\textrm{SNe}}(z)$ reconstruction of the expansion history from SNe Ia and anchors them with the gravitational lensing time delay angular diameter distances to determine $H_0$ with great accuracy. Further details will be provided in the following~sections.

\subsection{Time-Delay Angular Diameter Distance-$D_{A,\Delta~t}^{\textrm{SGL}} (z_l,z_s)$}\label{subsec21}

The time delay between multiple images of strongly lensed quasars has been used to infer $H_0$ severely~\cite{Wong2019,Birrer2020tax,2020MNRAS4931725K,Pandey2019yic}. As photons travel along null geodesics and originate from a distance source, they take distinct optical paths and must pass through different gravitational potentials~\cite{1964MNRAS128307R,PhysRevLett13789,2010ARAA4887T}. In~general, the~lensing time delay between any two images is determined by the geometry of the universe and the gravitational field of the lensing galaxy, which is expressed through the following relation:
\begin{equation}
    \Delta \tau = \frac{(1+z_l)}{c}\frac{D_{A_l} D_{A_s}}{D_{A_{ls}}} \left[ \frac{1}{2}(\vec{\theta}-\vec{\beta})^2 - \Psi(\vec{\theta})  \right],
\end{equation}
where $\Delta \tau$ represents the time-delay; $\vec{\theta}$ and $\vec{\beta}$ are the angular positions of the image and the source, respectively; and~$\Psi$ is the lens effective gravitational potential. For~a two-image lens system ($A$ and $B$) with the SIS mass profile describing the lens mass, it is possible to obtain the following:
\begin{equation}\label{Def_DADT}
    \Delta t=\Delta \tau (A) - \Delta \tau (B) = \frac{(1+z_l)}{2c}\frac{D_{A_l}D_{A_s}}{D_{A_{ls}}}[\theta_{A}^{2} - \theta_{B}^{2}].
\end{equation}

Defining the quantity $(1+z_l) \frac{D_{A_l}D_{A_s}}{D_{A_{ls}}} \equiv D_{A,\Delta t}^{\textrm{SGL}}(z_l,z_s)$ as the time-delay angular diameter distance, or~simply the time-delay distance, one obtains:
\begin{equation}\label{DADt}
    D_{A,\Delta t}^{\textrm{SGL}}(z_l,z_s) \equiv (1+z_l) \frac{D_{A_l}D_{A_s}}{D_{A_{ls}}} = \frac{2c \Delta t}{(\theta_{A}^{2} - \theta_{B}^{2})},
\end{equation}
where $D_{A_l} \equiv D_A(z_l)$ is the angular diameter distance (ADD) from observer to lens, \mbox{$D_{A_s} \equiv D_A (z_s)$} is the ADD from the observer to the source, and~$D_{A_{ls}} \equiv D_A (z_l, z_s)$ is the ADD from the lens to~the source. 

The use of time delay of SGL systems presents a significant advantage: their properties remain unaffected by dust absorption or changes in the source. Moreover, the~foundational assumptions behind the Singular Isothermal Sphere (SIS) or the Singular Isothermal Ellipsoid (SIE) models are valuable in gravitational lensing studies. They must provide a sufficiently accurate first-order approximation of the mean properties of galaxies relevant to statistical lensing. For~the two-image time-delay lensing systems supported by Equation~(\ref{DADt}), one utilizes a dataset compiled by~\cite{2013MNRAS4311528B}. It comprises 12 data points based on the Singular Isothermal Sphere (SIS) mass profile. However, while the selection of the SIS/SIE model is essential, it is important to note that this is not the most efficient way to ensure an accurate mass profile. Research from~\cite{Hugo2002} has shown that the presence of background matter tends to increase the image separations produced by lensing galaxies, a~finding supported by ray-tracing simulations in Cold Dark Matter (CDM) models\endnote{This effect is relatively small.}. Another research work indicated that the richer environments of early-type galaxies may host a higher ratio of dwarf to giant galaxies than those found in the field~\cite{christlein2000}. However, in Ref. ~\cite{keeton2000} it has shown that this effect nearly counteracts the influence of background matter, making the distribution of image separations largely independent of the environment. Furthermore, it has been predicted that lenses in groups exhibit a mean image separation of approximately $0.2$ arcseconds smaller than those found in the field. According to~\cite{2012JCAP03016C}, all the aforementioned factors can significantly influence the separation of images, potentially affecting the estimation of $D_{A,\Delta t}^{\textrm{SGL}}(z_l, z_s)$ by as much as $20\%$.

In addition, this paper also considers seven well-studied strong gravitational lensing (SGL) systems that have precise time-delay measurements between the lensed images, where the dataset is released by the H0LiCOW collaboration\endnote{Available online \url{http://www.h0licow.org}.}. The~redshifts of both the lenses and the sources, and the time-delay distances and angular diameter distances to the lenses for these systems are presented in Tables~1 and 2 of~\cite{Wong2019}. It is important to note that the observational quantity $D_{A, \Delta t}^{\textrm{SGL}}(z_l,z_s)$ is determined solely by the lens model and is independent of the cosmological model~\cite{Kelly2023mgv}.

\subsection{The Unanchored Luminosity Distances-$\Theta^{\textrm{SNe}} (z)$}

As mentioned before, observations of Type Ia supernovae can also provide the unanchored luminosity distance $\Theta^{\textrm{SNe}} (z)$ (see details in~\cite{Renzi2020fnx}). These distances are derived from the apparent magnitude of SNe Ia using the following relation:
\begin{equation}\label{SNemb}
    m_b(z) = 5 \log_{10}[\Theta^{\textrm{SNe}}(z)] - 5a_B,
\end{equation}
where $m_b$ represents the apparent magnitude. The~parameter $a_B$ is the intercept of the SNe Ia magnitude-redshift relation. In~the low-redshift limit, it is approximately given by $\log{cz} - 0.2 m_{x}^{0} $, but~for an arbitrary expansion history and $z>0$, it is defined by Equation~(5) from Ref.~\cite{Riess2016jrr}. The~parameter $m_{x}^{0}$ refers to the maximum-light apparent x-band brightness of an SNe Ia at the time of B-band peak, which is corrected to the fiducial color and luminosity, and is measured from the set of SNe Ia independent of any absolute scale (luminosity or distance). Using a Hubble diagram that includes up to 281 SNe Ia with a light-curve fitter employed to determine individual values of $m_{x}^{0}$ (see Figure~8 in Ref.~\cite{Riess2016jrr}), along with the current acceleration $q_0=-0.55$ and prior deceleration $j_0=1$, both measured through high-redshift SNe Ia independent of CMB or BAO~\cite{Riess2006fw,SDSS2014iwm}, the~authors found $a_B = 0.71273 \pm 0.00176$ with the uncertainty in $q_0$ contributing $0.1\%$ uncertainty. Therefore, combining the peak SNe Ia magnitudes with the intercept of their Hubble diagram in the relation $m_{x,i}^{0} + 5a_B$ (or Equation~(\ref{SNemb})), it provides a measure of distance independent of the choice of light-curve fitter, fiducial source, and~measurement filter. However, it is important to stress that $a_B$ has minimal model dependence on the choice of expansion rate form\endnote{The SH0ES team~\cite{Riess2021jrx} recently reviewed these assumptions and found no signs of inconsistency with~\cite{Riess2016jrr}.}.

This paper uses the largest compiled dataset of SNe Ia, known as Pantheon$+$ \cite{Scolnic2021amr}\endnote{Such a sample has a significant increase compared with the original Pantheon sample, particularly at lower redshifts, and~the full data releases are publicly available online \url{https://pantheonplussh0es.github.io/},  accessed on January 2025.}, to~obtain the unanchored luminosity distances. This dataset contains 1701 light curves from 1550 different SNe Ia in a redshift range of $0.001 \leq z \leq 2.261$. For~this paper's purposes, it is necessary to transform the apparent magnitudes of that sample into a set of unanchored luminosity distances, taking into account the relationship defined in Equation~(\ref{SNemb}):
\begin{equation}\label{H0DL}
    \Theta^{\textrm{SNe}} (z) = 10^{(m_b(z) + 5a_B)/5} = 10^{m'_B/5}, 
\end{equation}
where $m'_b(z) \equiv m_b (z) + 5a_B$. To~estimate the $\Theta^{\textrm{SNe}}(z)$ uncertainties, including their correlations, the covariance matrix of the apparent magnitudes (statistics + systematics) and the $a_B$ error are taken into consideration \endnote{If it is known that the covariance matrix is not diagonal, but~we decide to set the off-diagonal elements to zero, it will provide the fitter inaccurate uncertainty estimates. It could alter the conclusions drawn from the fit. Ignoring correlations may lead to underestimating the analysis's precision or affect the best-fit parameters' central values.}. It provides a more comprehensive statistical treatment of the data and improves the robustness of the resulting $H_0$ constraint. Thus, the~covariance matrix of $m'_b$ is as follows:
\begin{equation}
    \textbf{Cov} (\bm m'_b)=\textbf{Cov}(\bm m_b)+(5\sigma_{a{_B}})^2\bm I,
\end{equation}
where {$\bf I$} is the $n$-order square matrix, whose components are all equal to $1$ and $n=1701$. Note that all quantities in bold represent vectors or~matrices.

The covariance of the luminosity distances is calculated using the matrix transformation relation:
\begin{equation}\label{error_unanchored_luminosity}
\textbf{Cov}(\bm{ \Theta }^{SNe}) =
\left(\frac{\partial  \bm{  \Theta }^{SNe}} {\partial \bm{m'_b}}\right)\textbf{Cov}(\bm{m'_b}) \left(\frac{\partial  \bm{  \Theta   }^{SNe} }{\partial \bm{m'_b}}\right)^T,
\end{equation}
where $\frac{\partial  {\bm \Theta} ^{\rm SNe} }{\partial {\bm m'_b}}$ is the partial derivative matrix of the unanchored luminosity distance vector ${\bm \Theta}^{\rm SNe}$ regarding the vector ${\bm m'_b}$. Furthermore, it is well known that errors in redshift measurements for SNe Ia are negligible. Thus, no error bars are assigned to the variable $z$ in this paper, so that it is continuously varied in all Gaussian Processes conducted across the entire sample data~\cite{Colaco2023gzy}.

This paper utilizes the 2.7 Python Machine Learning GaPP\endnote{Available online \url{https://github.com/carlosandrepaes/GaPP}, accessed on December 2024.} code to perform the Gaussian Process (GP) regression~\cite{2012JCAP0036S} on Type Ia supernovae. The~GaPP code is widely recognized for its effectiveness in machine learning tasks and for being user-friendly and powerful. The~trained network provided by the GaPP code aims to forecast the unanchored luminosity distance at distinct $z$. For~this purpose, a~prior mean function and a covariance function are selected, which quantifiy the correlation between the dependent variable values of the reconstruction and are characterized by a set of hyperparameters~\cite{Seikel2013fda}. 
 In~general, zero is chosen as the prior mean function to prevent biased results, and it employs a Gaussian kernel as the covariance between two data points separated by a redshift distance of $z - z'$, which is given by the following relation:
\begin{equation}
\label{gaussian_kernel}
k(z,z')=\sigma^2 \exp\left(-\frac{(z-z')^2}{2l^2}\right),
\end{equation}
where $\sigma$ and $l$ represent the hyperparameters related to the variation of the estimated function and its smoothing scale, respectively. From~a Bayesian perspective, optimizing these hyperparameters typically provides a good approximation and can be computed much faster than other methods. Therefore, for~the unanchored luminosity distances, one maximizes the logarithm of the marginal likelihood
\begin{eqnarray}\label{log_likelihood}
    \ln {\mathcal{L}_{\textrm{SNe}}}&=&-\frac{1}{2}\bm \Theta^{\rm SNe} [\bm k(\bm z,\bm z)+\bm C_{\rm SNe}]^{-1}(\bm \Theta^{\rm SNe})^T \nonumber \\
&&   -\frac{1}{2}\ln |\bm k(\bm z,\bm z)+\bm C_{\rm SNe}|-\frac{n_d}{2}\ln 2\pi,
\end{eqnarray}
where $\bm{z}$ is the vector of redshift measurements of the Pantheon$+$ data, ${\bm k}({\bm z},{\bm z})$ is the covariance matrix used to describe the data as a GP and its elements are calculated with Equation~(\ref{gaussian_kernel}), ${\bm C_{\rm SNe}}$ is the covariance matrix of the data obtained by Equation~(\ref{error_unanchored_luminosity}), and $n_d$ is the number of data~points.

\begin{figure}[H]
\includegraphics[width=0.5\textwidth]{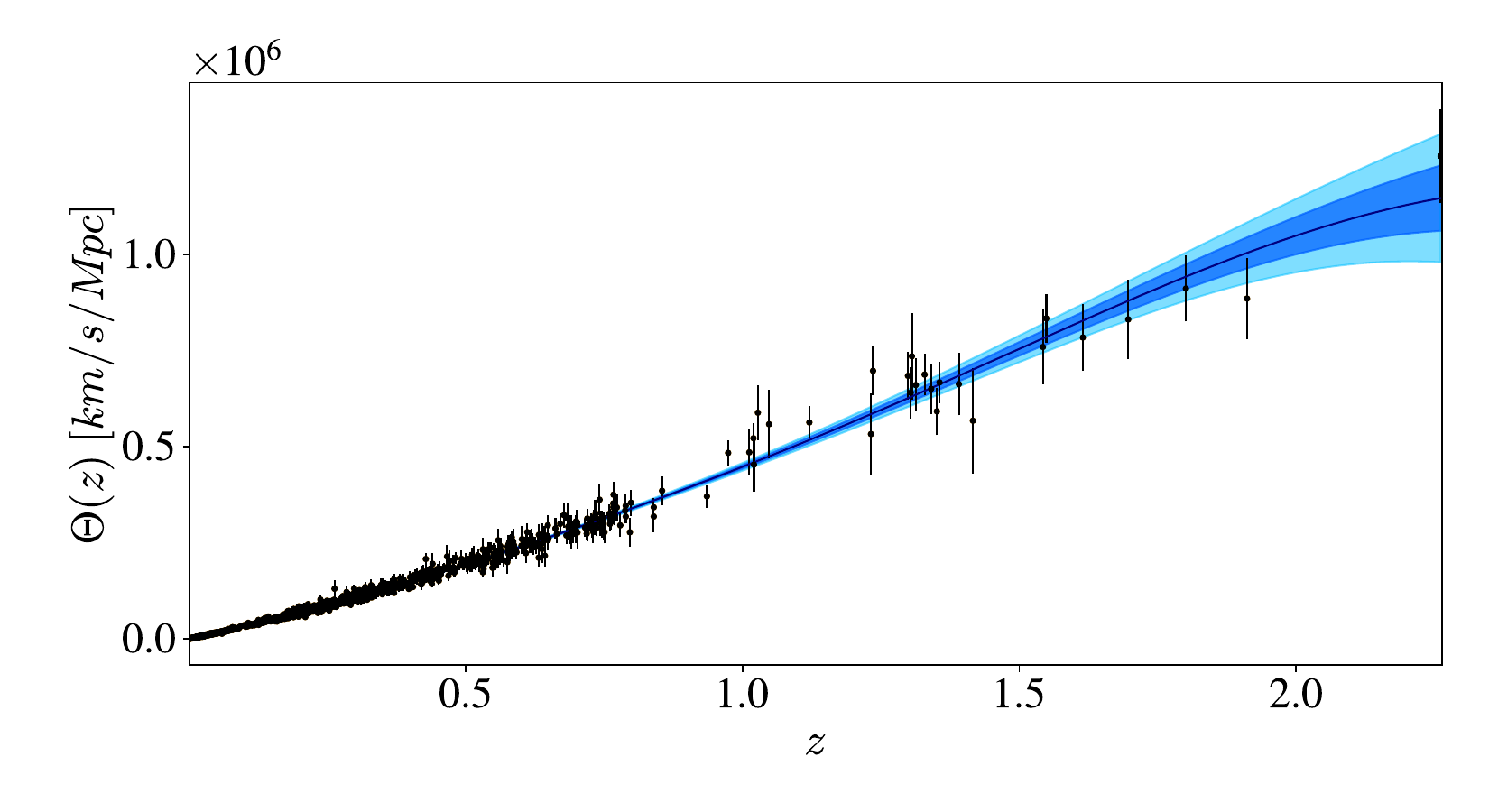}
\includegraphics[width=0.5\textwidth]{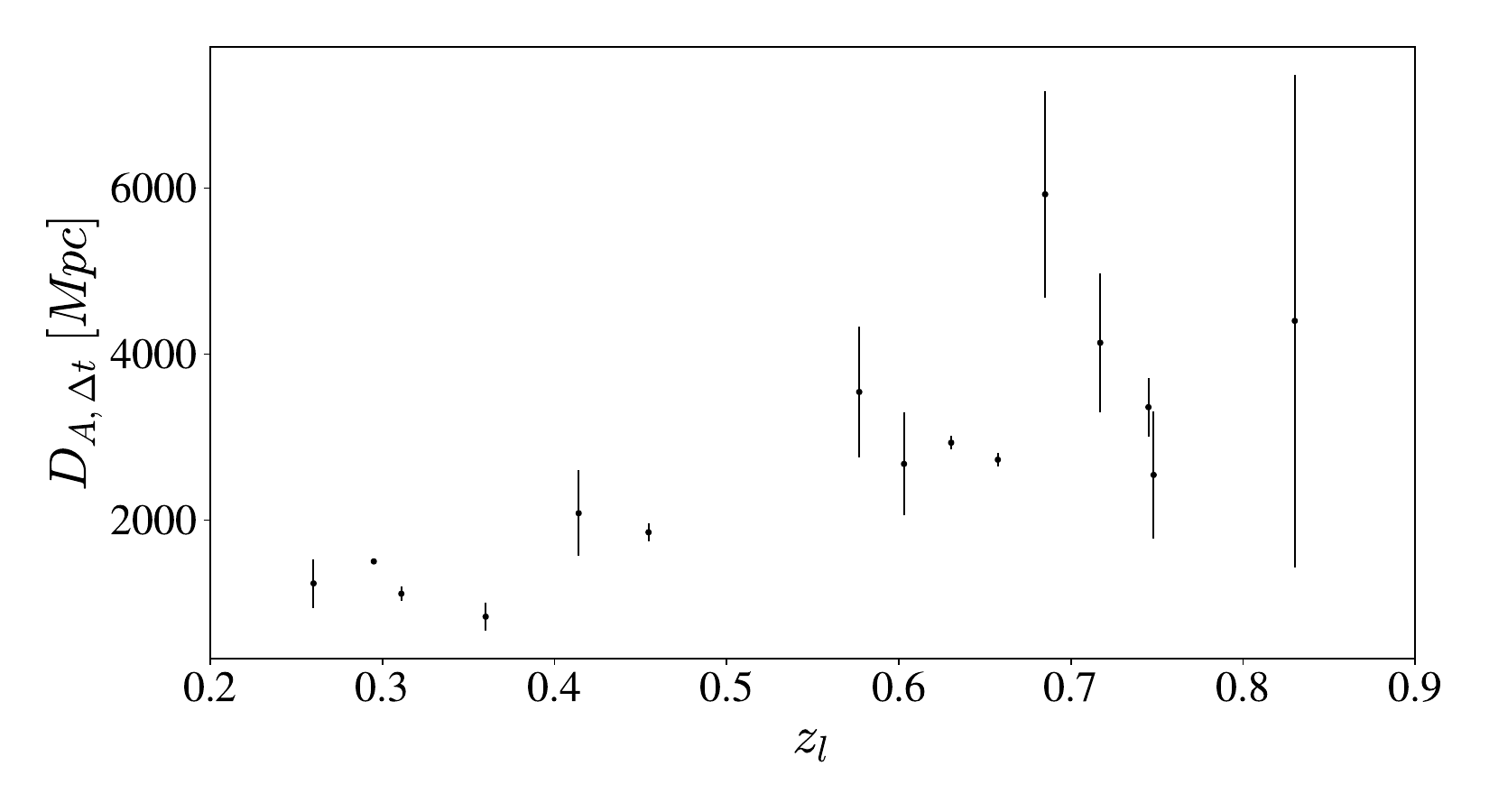}
\caption{(\textbf{Left}): The GP reconstruction of $\Theta (z) \equiv H_0 D_L(z) $ using the SNe Ia Pantheon$+$ compilation~\cite{Scolnic2021amr}. (\textbf{Right}): The 15 selected $D_{A,\Delta t}^{\textrm{SGL}}(z_l,z_s)$ data points according to $z_l$ from~\cite{2013MNRAS4311528B,Wong2019}. }
\label{data}
\end{figure}

As shown on the left panel of Figure~\ref{data}, the~reconstruction of SNe Ia using GP shows strong results at low redshifts. However, as~the redshift increases, the~uncertainties rise significantly due to the poor quality of data in that region. It is important to note that most kernels discussed in the literature tend to agree within the uncertainties of their predicted mean values~\cite{2012PhRvD85l3530S,Jesus2019nnk,Yang2015tzc,Li2021onq,Jesus2021bxq}. Thus, more data could improve future analyses, particularly SNe Ia observations at higher redshifts, as higher redshift reconstructions currently exhibit much more significant~errors.
\vspace{-10pt} 

\section{Results}\label{rasults}

In order to determine $H_0$, we consider a spatially flat universe to obtain the angular diameter distance between the lens and the source: $D_{A_{ls}} = D_{A_s} - \frac{1+z_l}{1+z_s}D_{A_l}$~\cite{Liao2020zko,Li2024103,Liao2019qoc}, where $D_{A_l} \equiv D_{A}(z_l)$, $D_{A_s} \equiv D_A (z_s)$, and $D_{A_{ls}} \equiv D_{A} (z_l, z_s)$. Combining this with \mbox{Equations \eqref{DLNA} and \eqref{DADt}}, it is possible to obtain the expression of unanchored time-delay distance ($H_0D_{A,\Delta t}^{\textrm{SNe}} (z_l, z_s) \equiv R ^{\textrm{SNe}}(z_l,z_s)$) based on the pair of unanchored luminosity distances from SNe Ia, given by the following:
\begin{equation}\label{h0da}
    R^{\textrm{SNe}} (z_l, z_s) = \frac{ \Theta^{\textrm{SNe}}(z_l)}{(1+z_l) - (1+z_s)\frac{ \Theta^{\textrm{SNe}}(z_l)}{ \Theta^{\textrm{SNe}} (z_s)}}.
\end{equation}
From Equation~(\ref{h0da}), it is possible to calculate/simulate relative time delay distances based on SNe Ia observations to match each time-delay distance measurement ($D_{A,\Delta t}^{\textrm{SGL}} (z_l,z_s)$) presented in Section \ref{subsec21}. However, not all systems have the corresponding pair of unanchored luminosity distances. This is owing to the fact that the source redshift of some lens systems exceeds the redshift of the SNe Ia sample. Excluding these systems, only 15 out of 19 time-delay distances are evaluated here, with~redshift ranges $0.26 \leq zl \leq 0.83$ and $0.654 \leq z_s \leq 2.033$ (see Right Panel of Figure~\ref{data}). 

Markov Chain Monte Carlo (MCMC) methods are used in this paper to estimate the posterior probability distribution function (pdf) of the free parameter supported by the \texttt{emcee} MCMC sampler~\cite{2013PASP125306F}. To~perform the plot, the~\texttt{GetDist} Python package~\cite{Lewis2019xzd} is required. Thus, the~following $\chi^2$ function is evaluated:
\begin{equation}
    \chi^2 = \Bigg( \frac{\bm{ R ^{\textrm{SNe}} (z_l, z_s)}}{\bm H_0}  - {\bm{D_{A,\Delta t}^{\textrm{SGL}}(z_l,z_s)}} \Bigg)\bm{C_{H_0}^{-1}} \Bigg( \frac{\bm{  R^{\textrm{SNe}} (z_l, z_s)}}{\bm H_0}  - {\bm{D_{A,\Delta t}^{\textrm{SGL}}(z_l,z_s)}} \Bigg)^T,
\end{equation}
where $\bm{R^{\textrm{SNe}} (z_l, z_s)}$ and $\bm{D_{A,\Delta t}^{\textrm{SGL}}(z_l,z_s)}$ are the data vectors supported by Equations \eqref{h0da} and \eqref{DADt}, respectively, and~${\bm H_0}$ is the free parameter. The~quantity $\bm C_{H_{0}}^{-1}$ is the inverse of the covariance matrix and it is given by $\bm{C_{H_0}} = \bm{C_{R^{\textrm{SNe}}}} + \bm{C_{D_{A,\Delta t}^{\textrm{SGL}}}}$, where $\bm{C_{D_{A,\Delta t}^{\textrm{SGL}}}}$ is the diagonal matrix\endnote{The $D_{A,\Delta t}^{\textrm{SGL}}(zl,zs)$ uncertainties are not correlated.}  of $D_{A,\Delta t}^{\textrm{SGL}} (z_l, z_s)$ and $C_{R^{\textrm{SNe}}}$ is obtained by using the matrix transformation relation regarding $z_l$ and $z_s$. 

The pdf is proportional to the product between the likelihood and prior ($P(H_0)$), that is,
\begin{equation}
    P(\bm{H_0}|\textrm{\bf{Data}}) \propto \mathcal{L} (\textrm{\bf{Data}}|\bm{H_0}) \times P(\bm{H_0}).
\end{equation}
In this analysis, one assumes a flat prior: \mbox{$H_0 = [50,100]$ km/s/Mpc}.

\begin{figure}[H]
\centering
\includegraphics[width=0.6\textwidth]{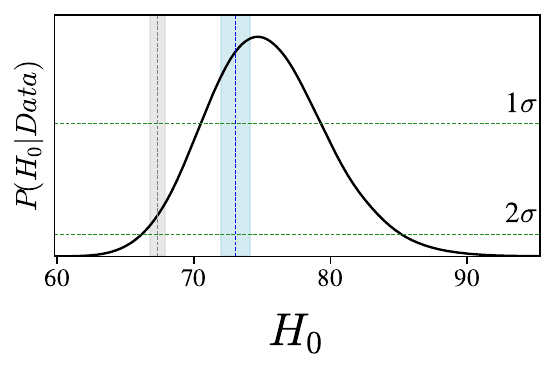}
\caption{The posterior probability distribution function for the free parameter $H_0$, with~a best-fit value of $75.57 \pm 4.415$ km/s/Mpc at the $1\sigma$ confidence level. The~grey and blue vertical dashed lines represent the estimates from Planck~\cite{Planck2020} and Riess~\cite{Riess2021jrx}, respectively, along with their corresponding $1\sigma$ confidence regions. The~light green horizontal dashed lines indicate the $1\sigma$ and $2\sigma$ confidence~levels.}
\label{Statistics}
\end{figure}

It is possible to obtain the following result at $68\%$ c.l. (see Figure~\ref{Statistics}): {$H_0 = 75.57 \pm 4.415$~km/s/Mpc}, with~an additional uncertainty $\sigma_{\text{int}} \approx 15\%$. This additional error accounts for potential random deviations from the simple isothermal sphere model and is necessary to achieve a reduced chi-squared value of approximately $1$. For~comparison, the~best-fit estimates from Riess {et~al}., 2020 (blue dashed vertical line) \cite{Riess2021jrx}, and Planck Collaboration, 2018 (grey dashed vertical line) \cite{Planck2020}, on $H_0$ with their respective $1\sigma$ regions are also shown in the Figure~\ref{Statistics}. As~may be seen, the~results presented here indicate good agreement with late-universe observations within $1\sigma$ c.l., and~the early-universe has a marginal agreement within the $2\sigma$ c.l. of the estimates. As the next generation of telescopes improves the size and precision of the available datasets, the methodology outlined here will greatly enhance the constraints on the Hubble constant in a model-independent manner.

\unskip

\section{Conclusions}\label{conclusions}

This paper presents a method for calculating the Hubble constant ($H_0$) without relying on any specific cosmological model. The~approach focuses solely on the validity of the cosmic distance duality relation, which indicates that if one has measurements of the unanchored luminosity distance ($\Theta(z)^{\textrm{SNe}}$) and the angular diameter distance ($D_A(z)$) at the same redshift ($z$), it is possible to obtain the value of $H_0$. It utilizes observations of Type Ia supernovae from Pantheon$+$ compilation to obtain $\Theta^{\textrm{SNe}}(z)$ and anchor with time-delay angular diameter distance measurements from Strong Gravitational Lensing. The~analysis yielded $H_0 = 75.57 \pm 4.415$ km/s/Mpc at a 68\% confidence level. As~shown in Figure~\ref{Statistics}, the~results presented here align well with the local estimate from Riess {et~al}. (2022) within the $1\sigma$ confidence region, indicating consistency with late-universe probes. Compared with the Planck Collaboration estimate, it has a marginal agreement within $2\sigma$ c.l. of this paper's estimate. This shows that a random error could flop either way, not always in the direction Planck ``wants''.

The next generation of telescopes, including the Euclid mission, the Nancy Grace Roman Space Telescope, and the Vera C. Rubin Observatory, will significantly increase the size of the available datasets. The~James Webb Space Telescope will also contribute to these efforts~\cite{Treu2022aqp}. As a result, the statistical and systematic uncertainties in strong gravitational lensing systems will decrease. Additionally, more observations of Type Ia supernovae at higher reshifts will become available, allowing the methodology outlined here to enhance the constraints on the Hubble constant in a model-independent manner. 

\vspace{6pt} 

\funding{This research was funded by Conselho Nacional de Desenvolvimento CientÍfico e Tecnologico (CNPq, National Council for Scientific and Technological Development) grant number 169625/2023-0.}

\dataavailability{The author confirms that the data supporting the findings of this study are available within the articles~\cite{Birrer2020tax, 2013MNRAS4311528B,Scolnic2021amr} and their Supplementary Materials.}

\acknowledgments{LRC thanks to Javier E. Gonzalez for his valuable contribution to this manuscript and the referees for their contributions that have greatly enhanced the overall quality of the work.}

\conflictsofinterest{The author declares no conflicts of interest.} 

\begin{adjustwidth}{-\extralength}{0cm}
\printendnotes[custom] 
\reftitle{References}

\isAPAandChicago{}{%
}
\PublishersNote{}
\end{adjustwidth}

\begin{thebibliography}{999}


\bibitem{Perlmutter1999}{Perlmutter, S.; Aldering, G.; Goldhaber, G.; Knop, R.A.; Nugent, P.; Castro, P.G.; Deustua, S.; Fabbro, S.; Goobar, A.; Groom, D.E.; {et~al}. Measurements of $\Omega$ and $\Lambda$ from 42 High-Redshift Supernovae. \textit{ Astrophys. J.} \textbf{1999}, \emph{517}, 565–586. }

\bibitem{Riess1998}{Riess, A.G.; Filippenko, A.V.; Challis, P.; Clocchiattia, A.; Diercks, A.; Garnavich, P.M.; Gilliland, R.L.; Hogan, C.J.; Jha, S.; Kirshner, R.P.; {et~al}. Observational Evidence from Supernovae for an Accelerating Universe and a Cosmological Constant. \textit{ Astron. J.} \textbf{1998}, \emph{116}, 1009–1038.}

\bibitem{Riess2016jrr}{Riess, A.G.; Filippenko, A.V.; Challis, P.; Clocchiattia, A.; Diercks, A.; Garnavich, P.M.; Gilliland, R.L.; Hogan, C.J.; Jha, S.; Kirshner, R.P.; {et~al}. A $2.4\%$ Determination of the Local Value of the Hubble Constant. \textit{ Astrophys. J.} \textbf{2016}, \emph{826}, 56.}

\bibitem{Riess2021jrx}{Riess, A.G.; Yuan, W.; Macri, L.M.; Scolnic, D.; Brout, D.; Casertano, S.; Jones, D.O.; Murakami, Y.; Brueval, L.; Brink, T.G.; {et~al}. A Comprehensive Measurement of the Local Value of the Hubble Constant with 1 km s$^{-1}$ Mpc$^{-1}$ Uncertainty from the Hubble Space Telescope and the SH0ES Team. \textit{ Astrophys. J. Lett.} \textbf{2022}, \emph{934},  L7.}

\bibitem{Reid2019}{Reid, M.J.; Pesce, D.W.; Riess, A.G. An Improved Distance to NGC 4258 and Its Im-plications for the Hubble Constant. \textit{ Astrophys. J. Lett.} \textbf{2019}, \emph{886}, L27.}

\bibitem{Freedman2021}{Freedman, W.L. Measurements of the Hubble Constant: Tensions in Perspective. \textit{ Astrophys. J.} \textbf{2021}, \emph{919}, 16.}

\bibitem{ScolnicL31}{Scolnic, D.; Riess, A.G.; Wu, J.; Li, S.; Anand, G.S.; Beaton, R.; Casertano, S.; Anderson, R.; Dhawan, S.; Ke, X. CATS: The Hubble Constant from Standardized TRGB and Type Ia Su-pernova Measurements. \textit{ Astrophys. J. Lett.} \textbf{2023}, \emph{954}, L31.}

\bibitem{Abbott2018}{Abbott, B.P.; Abbott, R.; Abbott, T.D.; Abraham, S.; Acernese, F.; Ackley, K.; Adams, C.; Adhikari, R.X.; Adya, V.B.; Affeldt, C.; {et~al}. A Gravitational-wave Measurement of the Hubble Constant Following the Second Observing Run of Advanced LIGO and Virgo. \textit{ Astro-Phys. J.} \textbf{2021}, \emph{909}, 218.}

\bibitem{Pogosian2020}{Pogosian, L.; Zhao, G.-B.; Jedamzik, K. Recombination-independent Determina-tion of the Sound Horizon and the Hubble Constant from BAO. \textit{Astrophys. J. Lett.} \textbf{2020}, \emph{904}, L17.}

\bibitem{Alam2021}{eBOSS Collaboration; Alam, S.; Aubert, M.; Avila, S.; Balland, C.; Bautista, J.E.; Bershady, M.A.; Bizyaev, D.; Blanton, M.R.; Bolton, A.S.; Bovy, J.; {et~al}. Completed SDSS-IV extended Baryon Oscillation Spectroscopic Survey: Cosmological implications from two decades of spectroscopic surveys at the Apache Point Observatory. \textit{Phys. Rev. D} \textbf{2021}, \emph{103}, 083533.}

\bibitem{Planck2020}{Planck Collaboration; Aghanim, N.; Akrami, Y.; Ashdown, M.; Aumont, J.; Bac-cigalupi, C.; Ballardini, M.; Banday, A.J.; Barreiro, R.B.; Bartolo, N.; Besak, S.; {et~al}. Planck 2018 results. VI. Cosmological parameters. \textit{Astron. Astrophys. } \textbf{2020}, \emph{641}, A6.}

\bibitem{Perivolaropoulos2024yxv}{Perivolaropoulos, Leandros. Hubble tension or distance ladder crisis? \textit{Phys. Rev. D} \textbf{2024}, \emph{110}, 123518.}

\bibitem{Abdalla2022yfr}{Abdalla, E.; Abellan, G.F.; Aboubrahim, A.; Agnello, A.; Akarsu, O.; Akrami, Y.; Alestas, G.; Aloni, D.; Amendola, L.; Anchordoqui, L.A.; {et~al}. Cosmology intertwined: A review of the particle physics, astrophysics, and cosmology associated with the cosmological tensions and anomalies. \textit{J. High Energy Astrophys.} \textbf{2022}, \emph{34}, 49--211.}

\bibitem{DiValentino2021izs}{Di Valentino, E.; Mena, O.; Pan, S.; Visinelli, L.; Yang, W.; Melchiorri, A.; Mota, D.F.; Riess, A.G.; Silk, J. In the realm of the Hubble tension\textemdash{}A review of solutions. \textit{Class. Quantum Gravity} \textbf{2021}, \emph{38}, 153001.}

\bibitem{Hu2023jqc}{Hu, J.-P.; Wang, F.-Y. Hubble Tension: The Evidence of New Physics. \textit{Universe} \textbf{2023}, \emph{9}, 94.}

\bibitem{McGaugh}{McGaugh, S.S. Discord in Concordance Cosmology and Anomalously Massive Early Galaxies. \textit{Universe} \textbf{2024}, \emph{10}, 48.}


\bibitem{Krishnan2021dyb}{Krishnan, C.; Mohayaee, R.; Colg\'ain, E.\'O.; Sheikh-Jabbari, M.M.; Yin, L. Does Hubble tension signal a breakdown in FLRW cosmology? \textit{Class. Quantum Gravity} \textbf{2021}, \emph{38}, 184001.}

\bibitem{Beenakker2021vff}{Beenakker, W.; Venhoek, D. A structured analysis of Hubble tension. \emph{arXiv} \textbf{2021}, arXiv:2101.01372.}

\bibitem{Benisty2022psx}{Benisty, D.; Mifsud, J.; Levi Said, J.; Staicova, D. On the robustness of the constancy of the Supernova absolute magnitude: Non-parametric reconstruction \& Bayesian approaches. \textit{Phys. Dark Universe} \textbf{2023}, \emph{39}, 101160.}

\bibitem{Alestas2021luu}{Alestas, G.; Camarena, D.; Di Valentino, E.; Kazantzidis, L.; Marra, V.; Nesseris, S.; Perivolaropoulos, L. Late-transition versus smooth $H(z)$-deformation models for the resolution of the Hubble crisis. \textit{Phys. Rev. D} \textbf{2022}, \emph{105}, 063538.}

\bibitem{Perivolaropoulos2021bds}{Perivolaropoulos, L.; Skara, F. Hubble tension or a transition of the Cepheid SnIa calibrator parameters?. \textit{Phys. Rev. D} \textbf{2021}, \emph{104}, 123511.}

\bibitem{Mortsell2021nzg}{Mortsell, E.; Goobar, A.; Johansson, J.; Dhawan, S. Sensitivity of the Hubble Constant Determination to Cepheid Calibration. \textit{ Astrophys. J.} \textbf{2022}, \emph{933}, 2 212.}




\bibitem{1964MNRAS128307R}{Refsdal, S. On the possibility of determining Hubble's parameter and the masses of galaxies from the gravitational lens effect. \textit{Mon. Not. R. Astron. Soc.} \textbf{1964}, \emph{128}, 307.}

\bibitem{Birrer2022chj}{Birrer, S.; Millon, M.; Sluse, D.; Shajib, J.; Courbin, F.; Koopmans, L.V.E.; Suyu, S.H.; Treu, T. Time-Delay Cosmography: Measuring the Hubble Constant and Other Cosmological Param-eters with Strong Gravitational Lensing. \textit{Space Sci. Rev.} \textbf{2024}, \emph{220}, 48.}

\bibitem{Saha2024axf}{Saha, P.; Sluse, D.; Wagner, J.; Williams, L.L.R. Essentials of Strong Gravitational Lensing. \textit{Space Sci. Rev.} \textbf{2024}, \emph{220}, 12.}

\bibitem{2018MNRAS4811041T}{Treu, T.; Agnello, A.; Baumer, M.A.; Birrer, S.; Buckley-Geer, E.J.; Courbin, F.; Kim, Y.J.; Lim, H.; Marshall, P.J.; Nord, B.; {et~al}. The STRong lensing Insights into the Dark Energy Survey (STRIDES) 2016 follow-up campaign-I. Overview and classification of candidates selected by two techniques. \textit{Mon. Not. R. Astron. Soc.} \textbf{2018}, \emph{481}, 1041--1054.}

\bibitem{2023MNRAS5203305L}{Lemon, C.; Anguita, T.; Auger-Williams, M.W.; Courbin, F.; Galan, A.; McMahon, R.; Neira, F.; Oguri, M.; Schechter, P.; Shajib, A.;~et~al. Gravitationally lensed quasars in Gaia-IV. 150 new lenses, quasar pairs, and projected quasars. \textit{Mon. Not. R. Astron. Soc.} \textbf{2023}, \emph{520}, 3305--3328.}

\bibitem{Rodney2021keu}{Rodney, S.A.; Brammer, G.B.; Pierel, J.D.R.; Richard, J.; Toft, S.; O'Connor, K.F.; Akhshik, M.; Whitaker, K.E. A gravitationally lensed supernova with an observable two-decade time delay. \textit{Nat. Astron.} \textbf{2021}, \emph{5}, 1118--1125.}

\bibitem{Millon2020ugy}{Millon, M.; Courbin, F.; Bonvin, V.; Buckley-Geer, E.; Fassnacht, C.D.; Frie-man, J.; Marshal, P.J.; Suyu, S.H.; Treu, T.; Anguita, T.; {et~al}. TDCOSMO-II. Six new time delays in lensed quasars from high-cadence monitoring at the MPIA 2.2 m telescope. \textit{Astron. Astrophys.} \textbf{2020}, \emph{642}, A193.}

\bibitem{2017arXiv170609424C}{Courbin, F.; Bonvin, V.; Buckley-Geer, E.; Fassnacht, C.D.; Frieman, J.; Lin, H.; Marshall, P.J.; Suyu, S.H.; Treu, T.; Anguita, T.; {et~al}. COSMOGRAIL XVI: Time delays for the quadruply imaged quasar DES J0408-5354 with high-cadence photometric monitoring. \textit{Astron. Astrophys.} \textbf{2018}, \emph{609}, A71.}

\bibitem{Wong2019}{Wong, K.C.; Suyu, S.H.; Chen, G.C.-F.; Rusu, C.E.; Millon, M.; Sluse, D.; Bonvin, V.; Fassnacht, C.D.; Taubenberger, S.; Auger, M.W.; {et~al}. H0LiCOW–XIII. A 2.4 per cent measurement of $H_0$ from lensed quasars: 5.3$\sigma$ tension between early- and late-Universe probes. \textit{Mon. Not. R. Astron. Soc.} \textbf{2019}, \emph{498}, 1420--1439.}

\bibitem{Millon2019slk}{Millon, M.; Galan, A.; Courbin, F.; Treu, T.; Suyu, S.H.; Birrer, S.; Chen, G.C.-F.; Shajib, A.J.; Sluse, D.; Wong, K.C.; {et~al}. TDCOSMO. I. An exploration of systematic uncertainties in the inference of $H_0$ from time-delay cosmography. \textit{Astron. Astrophys.} \textbf{2020}, \emph{639}, A101.}

\bibitem{Birrer2020tax}{Birrer, S.; Shajib, A.J.; Galan, A.; Millom, M.; Treu, T.; Agnello, A.; Auger, M.; Chen, G.C.-F.; Christensen, L.; Collett, T.; {et~al}. TDCOSMO-IV. Hierarchical time-delay cosmography\textendash{}joint inference of the Hubble constant and galaxy density profiles. \textit{Astron.  Astrophys.} \textbf{2020}, \emph{643}, A165.}



\bibitem{Aubourg15}{Aubourg, É.; Bailey, S.; Bautista, J.E.; Beutler, F.; Bhardwaj, V.; Bizyaev, D.; Blanton, M.; Blomqvist, M.; Bolton, A.S.; Bovy, J.; Brewington, H.; {et~al}. Cosmological implications of baryon acoustic oscillation measurements. \textit{Phys. Rev. D} \textbf{2015}, \emph{92}, 123516 .}

\bibitem{Cuesta3463}{Cuesta, A.J.; Verde, L.; Riess, A.; Jimenez, R. Calibrating the cosmic distance scale ladder: The role of the sound-horizon scale and the local expansion rate as distance anchors. \textit{Mon. Not. R. Astron. Soc.} \textbf{2015}, \emph{448}, 3463--3471.}

\bibitem{Liao2020zko}{Liao, K.; Shafieloo, A.; Keeley, R.E.; Linder, E.V. Determining Mod-el-independent H 0 and Consistency Tests. \textit{Astrophys. J. Lett.} \textbf{2020}, \emph{895},  L29.}

\bibitem{Liao2019qoc}{Liao, K.; Shafieloo, A.; Keeley, R.E.; Linder, E.V. A model-independent deter-mination of the Hubble constant from lensed quasars and supernovae using Gaussian process regression. \textit{Astrophys. J. Lett.} \textbf{2019}, \emph{886},  L23.}

\bibitem{Li2024elb}{Li, X.; Liao, K. Determining Cosmological-model-independent H$_{0}$ with Gravitationally Lensed Supernova Refsdal. \textit{ Astrophys. J.} \textbf{2024}, \emph{966}, 121.}

\bibitem{Li2024103}{Li, X.; Keeley, R.E.; Shafieloo, A.; Liao, K. A Model-independent Method to De-termine $H_0$ Using Time-delay Lensing, Quasars, and Type Ia Supernovae. \textit{ Astrophys. J.} \textbf{2024}, \emph{960}, 103.}

\bibitem{Gong2024yne}{Gong, X.; Liu, T.; Wang, J. Inverse distance ladder method for determining $H_0$ from angular diameter distances of time-delay lenses and supernova observations. \textit{ Eu-Ropean Phys. J. C} \textbf{2024}, \emph{84}, 873.} 

\bibitem{Colaco2023gzy}{Cola\c{c}o, L.R.; Ferreira, M.; Holanda, R.F.L.; Gonzalez, J.E.; Nunes, R.C. A Hubble constant estimate from galaxy cluster and type Ia SNe observations. \textit{J. Cosmol. Astropart. Phys.} \textbf{2024}, \emph{5}, 098.}

\bibitem{Gonzalez2024qjs}{Gonzalez, J.E.; Ferreira, M.; Cola\c{c}o, L.R.; Holanda, R.F.L. Unveiling the Hubble constant through galaxy cluster gas mass fractions. \textit{Phys. Lett. B} \textbf{2024}, \emph{857}, 138982.}

\bibitem{Camarena2019rmj}{Camarena, D.; Marra, V. A new method to build the (inverse) distance ladder. \textit{Mon. Not. R. Astron. Soc.} \textbf{2020}, \emph{495}, 2630--2644.}



\bibitem{Scolnic2021amr}{Scolnic, D.; Brout, D.; Carr, A.; Riess, A.G.; Davis, T.M.; Dwomoh, A.; Jones, D.O.; Ali, N.; Charvu, P.; Chen, R.; {et~al}. The Pantheon+ Analysis: The Full Data Set and Light-curve Release. \textit{ Astrophys. J.} \textbf{2022}, \emph{938}, 113.}

\bibitem{2013MNRAS4311528B}{Balm{\`e}s, I.; Corasaniti, P.S. Bayesian approach to gravitational lens model selection: Constraining H$_{0}$ with a selected sample of strong lenses. \textit{Mon. Not. R. Astron. Soc.} \textbf{2013}, \emph{431}, 1528--1540.}




\bibitem{Wang2024rxm}{Wang, M.; Fu, X.; Xu, B.; Huang, Y.; Yang, Y.; Lu, Z. Testing the cosmic dis-tance duality relation with Type Ia supernova and transverse BAO measurements. \textit{ Eur. Phys. J. C} \textbf{2024}, \emph{84}, 7.}

\bibitem{Favale2024sdq}{Favale, A.; G\'omez-Valent, A.; Migliaccio, M. 2D vs. 3D BAO: Quantification of their tension and test of the Etherington relation. \textit{Phys. Lett. B} \textbf{2024}, \emph{858}, 139027.}

\bibitem{Kumar2021djt}{Kumar, D.; Rana, A.; Jain, D.; Mahajan, S.; Mukherjee, A.; Holanda, R.F.L. A non-parametric test of variability of Type Ia supernovae luminosity and CDDR. \textit{J. Cos-Mology Astropart. Phys.} \textbf{2022}, \emph{1}, 1.}

\bibitem{Lima2021slf}{Lima, F.S.; Holanda, R.F.L.; Pereira, S.H.; da Silva, W.J.C. On the cosmic distance duality relation and strong gravitational lens power-law density profile. \textit{J. Cosmol. Astropart. Phys.} \textbf{2021}, \emph{8}, 035.}

\bibitem{Li2023frp}{Li, P. Distance Duality Test: The Evolution of Radio Sources Mimics a Non-expanding Universe. \textit{Astrophys. J. Lett.} \textbf{2023}, \emph{950},  L14.}

\bibitem{Jesus2024nrl}{Jesus, J.F.; Gomes, M.J.S.; Holanda, R.F.L.; Nunes, R.C. High-redshift cosmography with a possible cosmic distance duality relation violation. \textit{J. Cosmol. Astropart. Phys.} \textbf{2025}, \emph{1}, 88.}

\bibitem{Ellis2007}{Ellis, G.F.R. On the definition of distance in general relativity: I. M. H. Etherington (Philosophical Magazine ser. 7, vol. 15, 761 (1933)). \textit{Gen. Relativ. Gravit.} \textbf{2007}, \emph{39}, 7.}

\bibitem{CDDR}{Etherington, I.M.H. On the definition of distance in general relativity. \textit{Phil. Mag. ser.} \textbf{1933}, \emph{15}, 7 761.}

\bibitem{Renzi2020fnx}{Renzi, F.; Silvestri, A. Hubble speed from first principles. \textit{Phys. Rev. D} \textbf{2023}, \emph{107}, 2.}

\bibitem{2020MNRAS4931725K}{Kochanek, C.S. Overconstrained gravitational lens models and the Hubble constant. \textit{Mon. Not. R. Astron. Soc.} \textbf{2020}, \emph{493}, 2.}

\bibitem{Pandey2019yic}{Pandey, S.; Raveri, M.; Jain, B. Model independent comparison of supernova and strong lensing cosmography: Implications for the Hubble constant tension. \textit{Phys. Rev. D} \textbf{2020}, \emph{102}, 2.}

\bibitem{PhysRevLett13789}{Shapiro, I.I. Fourth Test of General Relativity. \textit{Phys. Rev. Let-Ters} \textbf{1964}, \emph{13}, 26.}

\bibitem{2010ARAA4887T}{Treu, T. Strong Lensing by Galaxies. \textit{ Annu. Rev. Astron-Omy Astrophys.} \textbf{2010}, \emph{48}, 87--125.}

\bibitem{Hugo2002}{Martel, H.; Premadi, P.; Matzner, R. Light Propagation in Inhomogeneous Universes. III. Distributions of Image Separations. \textit{ Astrophys. J.} \textbf{2002}, \emph{570}, 17.}

\bibitem{christlein2000}{Christlein, D. The Dependence of the Galaxy Luminosity Function on Enviroment. \textit{ Astrophys. J.} \textbf{2000}, \emph{545}, 145.}

\bibitem{keeton2000}{Keeton, C.R.; Christlein, D.; Zabludoff, A.I. What Fraction of Gravitational Lens Galaxies Lie in Groups? \textit{ Astrophys. J.} \textbf{2000}, \emph{545}, 129.}

\bibitem{2012JCAP03016C}{Cao, S.; Pan, Y.; Biesiada, M.; Godlowski, W.; Zhu, Z.-H. Constraints on cosmological models from strong gravitational lensing systems. \textit{J. Cosmol. Astro-Part. Phys.} \textbf{2012}, \emph{2012}, 3.}

\bibitem{Kelly2023mgv}{Kelly, P.L.; Rodney, S.; Treu, T.; Oguri, M.; Chen, W.; Zitrin, A.; Birrer, S.; Bonvin, V.; Dessart, L.; Diego, J.M.; {et~al}. Constraints on the Hubble constant from supernova Refsdal\textquoteright{}s reappearance. \textit{Science} \textbf{2023}, \emph{380},  abh1322.}

\bibitem{Riess2006fw}{Riess, A.G.; Strolger, L.-G.; Casertano, S.; Ferguson, H.C.; Mobasher, B.; Gold, B.; Challis, P.J.; Filippenko, A.V.; Jha, S.; Li, W.;  {et~al}. New Hubble Space Telescope Discoveries of Type Ia Supernovae at z\ensuremath{>}=1: Narrowing Constraints on the Early Behavior of Dark Energy. \textit{ Astrophys. J.} \textbf{2007}, \emph{659}, 98--121.}

\bibitem{SDSS2014iwm}{Betoule, M.; Kessler, R.; Guy, J.; Mosher, J.; Hardin, D.; Biswas, R.; Astier, P.; El-Hage, P.; Koing, M.; Kuhlmann, S.; {et~al}. Improved cosmological constraints from a joint analysis of the SDSS-II and SNLS supernova samples. \textit{Astron. Astrophys.} \textbf{2014}, \emph{568}, A22.}

\bibitem{2012JCAP0036S}{Seikel, M.; Clarkson, C.; Smith, M. Reconstruction of dark energy and expan-sion dynamics using Gaussian processes. \textit{J. Cosmol. Astropart. Phys.} \textbf{2012}, \emph{2012}, 036.}

\bibitem{Seikel2013fda}{Seikel, M.; Clarkson, C. Optimising Gaussian processes for reconstructing dark energy dynamics from supernovae. \emph{arXiv} \textbf{2013}, arXiv:1311.6678.}

\bibitem{2012PhRvD85l3530S}{Shafieloo, A.; Kim, A.G.; Linder, E.V. Gaussian process cosmography. \textit{Phys. Rev. D} \textbf{2012}, \emph{85}, 123530.}


\bibitem{Jesus2019nnk}{Jesus, J.F.; Valentim, R.; Escobal, A.A.; Pereira, S.H. Gaussian Process Estima-tion of Transition Redshift. \textit{J. Cosmol. Astropart. Phys.} \textbf{2020}, \emph{4}, 053.}

\bibitem{Yang2015tzc}{Yang, T.; Guo, Z.-K.; Cai, R.-G. Reconstructing the interaction between dark en-ergy and dark matter using Gaussian Processes. \textit{Phys. Rev. D} \textbf{2015}, \emph{91}, 123533.}

\bibitem{Li2021onq}{Li, X.; Keeley, R.E.; Shafieloo, A.; Zheng, X.; Cao, S.; Biesiada, M.; Zhu, Z.-H. Hubble diagram at higher redshifts: Model independent calibration of quasars. \textit{Mon. Not. R. Astron. Soc.} \textbf{2021}, \emph{507}, 919--926.}

\bibitem{Jesus2021bxq}{Jesus, J.F.; Valentim, R.; Escobal, A.A.; Pereira, S.H.; Benndorf, D. Gaussian processes reconstruction of the dark energy potential. \textit{J. Cosmol. Astropart. Phys.} \textbf{2022}, \emph{11}, 037.}




\bibitem{2013PASP125306F}{Foreman-Mackey, D.; Hogg, D.W.; Lang, D.; Goodman, J. emcee: The MCMC Hammer. \textit{Publ. Astron. Soc. Pac.} \textbf{2013}, \emph{125}, 306.}

\bibitem{Lewis2019xzd}{Lewis, A. GetDist: A Python package for analysing Monte Carlo samples. \emph{arXiv} \textbf{2019}, arXiv:1910.13970.}



\bibitem{Treu2022aqp}{Treu, T.; Suyu, S.H.; Marshall, P.J. Strong lensing time-delay cosmography in the 2020s. \textit{ Astron. Astrophys. Rev.} \textbf{2022}, \emph{30}, 8.}




\end{thebibliography}
\end{document}